\documentclass[a4paper,twoside]{article}
\newcommand{\affil}[1]{$^{\rm #1}$}

\usepackage[authoryear]{natbib}
\bibpunct{(}{)}{;}{a}{}{,}
\usepackage{graphicx}
\date{} %Please leave the date blank
\def\reference{\parskip 0pt\par\noindent\hangindent 0.5 truecm}
\newcommand\ion[2]{#1$\;${\footnotesize{#2}}\relax}% 
\newcommand{\hst}{{\sl HST}}
\newcommand{\kms}{\mbox{km\,s$^{-1}$}}
\newcommand{\lya}{\mbox{Ly$\alpha$}}
\newcommand{\Ha}{\mbox{H$\alpha$}}
\title{\large\bf\flushleft A Direct Detection of Gas Accretion:  The Lyman Limit System in 3C~232}
\author{\parbox{\textwidth}{\flushleft
\vspace{-0.5cm}
{\em John T. Stocke\affil{a,b}, Brian A. Keeney\affil{a} and Charles W. Danforth\affil{a}}\\
\vspace{0.4cm}
{\small \affil{a}\, Center for Astrophysics \& Space Astronomy, University of Colorado, Boulder CO 80309 USA}\\
{\small \affil{b}\,Email: stocke@casa.colorado.edu}}}
\begin{document}
\twocolumn[
\begin{changemargin}{.8cm}{.5cm}
\begin{minipage}{.9\textwidth}
\vspace{-1cm}
\maketitle
%
%
%%%%%%%%%%%%%     ABSTRACT    %%%%%%%%%%%%%
%Abstract of no more than 200 words here.
\small{\bf Abstract:}
The gas added and removed from galaxies over cosmic time greatly affects their stellar populations and star formation rates. QSO absorption lines studies in close QSO/galaxy pairs create a unique opportunity to study the physical conditions and kinematics of this gas. Here we present new {\em Hubble Space Telescope} (\hst) images of the QSO/galaxy pair, 3C~232/NGC~3067. The quasar spectrum contains a Lyman-limit (N$_{\rm H\,I} = 1 \times 10^{20}~{\rm cm^{-2}}$) absorption system (LLS) at $cz =1421$~\kms\ that is associated with the nearby SAB galaxy NGC~3067. Previous work identifies this absorber as a high-velocity cloud (HVC) in NGC~3067 but the kinematics of the absorbing gas, infalling or outflowing, were uncertain. The \hst\ images presented here establish the orientation of NGC~3067 and so establish that the LLS/HVC is infalling. Using this system as a prototype, we extend these results to higher-$z$ \ion{Mg}{II}/LLS to suggest that \ion{Mg}{II}/LLSs are a sightline sampling of the so-called `cold mode accretion' (CMA) infalling onto luminous galaxies. But to match the observed \ion{Mg}{II} absorber statistics, the CMA must be more highly ionized at higher redshifts. The key observations needed to further the  study of low-$z$ LLSs is \hst/UV spectroscopy, for which a new instrument, the Cosmic Origins Spectrograph (COS), has just been installed greatly enhancing our observational capabilities.

%%%%%%%%%%%%%     KEYWORDS    %%%%%%%%%%%%%
 
% Please write all keywords in lower case. PASA uses the
% standard list of subject headings adopted by The Astrophysical Journal
% and available from http://www.journals.uchicago.edu/ApJ/keywords_text.html.
% Keywords are separated by em-dashes, i.e. ---

%%%%%%%%DO NOT EDIT%%%%%%%%%%%%
\medskip
\medskip
\end{minipage}
\end{changemargin}
]
\small
%%%%%%%%EDIT FROM HERE%%%%%%%%%%%%

\section{Introduction}
%Please see the PASA Style Guide for help with correct layout for your manuscript.
%Examples of tables and figures are given below.

Understanding the evolution of galaxies over cosmic time must include both the amounts and metallicities of gas infall and outflow in order to understand details of stellar evolution (e.g., the `G-dwarf problem') and star formation in galaxy disks as well as the direct observations of these gas flows around galaxies (starburst and AGN winds and infalling high velocity clouds).  Understanding the kinematics and constituency of this `circumgalactic medium' (CGM) is also critical for understanding the more remote `intergalactic medium' (IGM), wherein resides the bulk of the baryons ($> 80\%$) in the local Universe ($ z\leq 0.1$). Until recently our knowledge of gas flows around galaxies has been severely limited to that portion of the CGM (e.g., hot X-ray coronae and optical emission line filaments) that can be seen at high contrast in emission very near some galaxies.

But for several decades large-aperture ground-based optical telescopes have studied in great detail the plethora of absorption lines (several hundreds per unit redshift) in the spectra of distant QSOs and quasars. These absorption lines of \ion{H}{I} (in the Lyman lines) and light metals (mostly in strong resonance lines like \ion{C}{II} 1335~\AA, \ion{C}{III} 977~\AA, \ion{C}{IV} 1548,1551~\AA, \ion{Si}{III} 1206~\AA, \ion{Si}{IV} 1398,1402~\AA, and \ion{O}{VI} 1031,1038~\AA) have allowed us to characterize some of the more basic properties of the CGM and IGM from $z = 1.8$--6. However, optical detections of the strong near-UV \ion{Mg}{II} doublet (2796, 2803~\AA) at $z \geq 0.2$ in conjunction with spectroscopy of the far-UV transitions listed above have also begun to set strong constraints on the nature of the CGM. 
    
The advent of the {\em Hubble Space Telescope} (\hst) and its UV spectrographs has provided a new method for studying these CGM gas flows directly via absorption lines in the spectra of bright AGN targets close on the sky to relatively nearby galaxies. While previous UV spectrographs aboard \hst, including the recently revived Space Telescope Imaging Spectrograph (STIS), have made some progress on this front, the now successfully-deployed, tested, and calibrated Cosmic Origins Spectrograph (COS) will revolutionize studies of galaxy infall/outflow due to its 10--20 times greater sensitivity than STIS at comparable spectral resolution ($R \approx 18\,000$; Froning \& Green 2009). This greatly enhanced sensitivity is already allowing high signal-to-noise (${\rm SNR} \sim 20$--50) spectra to be obtained in only a few orbits for QSOs too faint to have been easily observed with STIS. 

While \hst's UV capability opens these studies to the most recent 75\% of a Hubble time ($z<1.8$), the greatest advantages for CGM and IGM studies is at the very lowest redshifts ($z<0.1$). These very nearby absorbers can provide bright, well-resolved prototypes (including absorption from our own Milky Way) whose detailed study yields the necessary baseline parameters for absorption systems already discovered out to nearly the highest redshifts where galaxies have been discovered. Studies of these nearest of all IGM \& CGM \lya\ and metal-line absorbers also profit from recent, large-angle galaxy surveys that are the most complete at the lowest redshifts. Additionally, a host of techniques can be employed at the lowest redshifts (e.g., \ion{H}{I} 21-cm emission line rotation curves and neutral gas distributions; detailed morphological studies of the absorbing galaxy and even the precise location of the sightline through the absorbing galaxy) which are not possible at higher redshifts.

Here we present new results which further illuminate the detailed properties of a very nearby ($cz =1421$~\kms) Lyman-limit system (LLS; the gas is optically-thick at the Lyman-limit; N$_{\rm H\,I}= 2\times10^{17}$ to $2\times10^{20}~{\rm cm^{-2}}$) due to the nearly edge-on spiral galaxy NGC~3067. This absorber is detected in the UV, optical and radio spectra of the bright quasar 3C~232 projected 11~kpc above the galaxy's nucleus. In a previous paper on this system (Keeney et~al. 2005) we showed that the detailed properties of the LLS are identical to the properties of Galactic high-velocity clouds (HVCs) and so had identified this absorber as an HVC near NGC~3067. Although the \ion{H}{I} 21-cm rotation curve for NGC~3067 strongly suggests that the kinematics of the LLS are {\bf infalling} onto the disk of the galaxy, the importance of this inference for understanding HVCs and accretion flows onto normal galaxies requires us to revisit this conclusion using new data. Newly obtained \hst/WFPC2 \Ha\ and continuum images of NGC~3067 allow us to determine the orientation of the galaxy disk in space and to solidify our previous conclusion that the LLS is CGM gas being accreted by NGC~3067. Using this system as a prototype for more distant LLSs found using strong \ion{Mg}{II} absorbers at higher redshifts, we then use this result to place observational constraints on the so-called `cold mode accretion' (CMA) seen to dominate gas accretion onto galaxies in current epoch numerical simulations. 

We begin this paper by introducing results on another very nearby QSO/galaxy pair PKS~1327--206/ ESO~1327--2041, which illustrates both the many advantages of studying very nearby absorption systems and also the limitations. In Section~3 we present the new \hst\ images of the 3C~232/NGC~3067 system and the direct inferences it allows for constraining the nature of this absorbing cloud and its status as an HVC in another galaxy. In Section~4 we discuss this result in the context of the significant dataset on \ion{Mg}{II}/LLSs amassed by ground-based studies.  Section~5 discusses the future of this work.

\section{PKS~1327--206/ESO~1327--2041: A Very Nearby Damped \lya\ System}

Figure~1 is a composite {\em B}-, {\em R}-, and {\em I}-band \hst/WFPC2 continuum image with ground-based \Ha\ from the Apache Point 3.5-m overlaid in red, showing the projection of a bright quasar ($z=1.169$) onto an outer spiral arm 14~kpc from the nucleus of a strongly interacting lenticular plus spiral galaxy system ($cz=5340$~\kms). In this paper we summarize some of our conclusions involving the high column density absorption systems detected in the optical/UV and radio spectra of the quasar in order to set the context for our 3C~232 discussion. But this is clearly an unusual system, a `polar-ring' galaxy. In their catalogue of similar systems, Whitmore et~al. (1990) find that only 0.5\% of nearby S0 galaxies exhibit polar ring structures like the one seen in Figure 1. A detailed discussion of this system can be found elsewhere (Keeney et~al. 2010).

\begin{figure}[!t]
\begin{center}
\includegraphics[scale=0.58, angle=0]{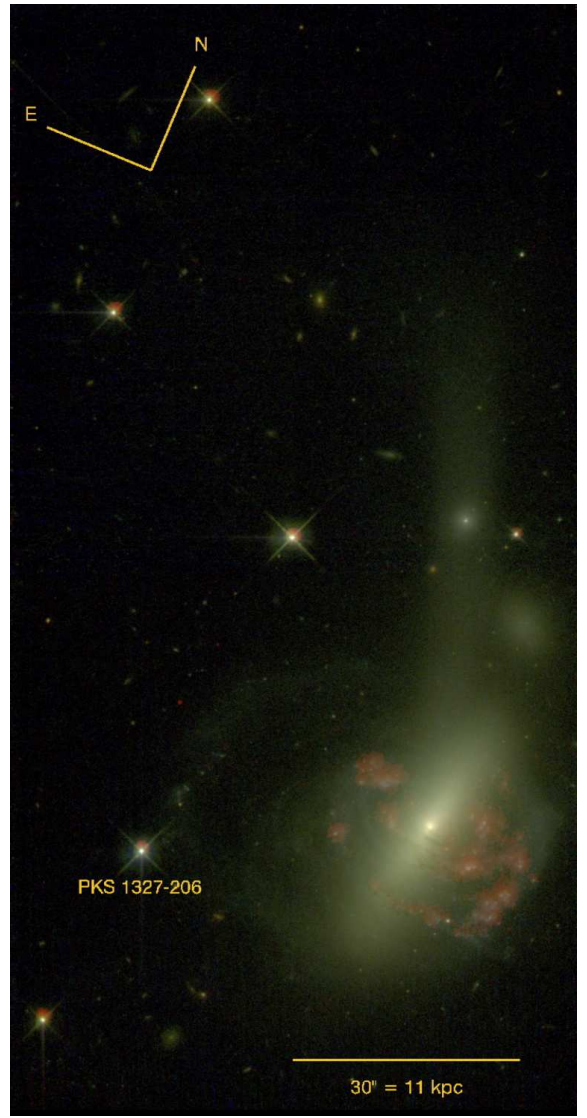}
\caption{\small A composite {\em B}-, {\em R}-, and {\em I}-band \hst/WFPC2 continuum image of the PKS~1327--206/ESO~1327--2041 system with ground-based \Ha\ from the Apache Point 3.5-m overlaid in red. Notice that the QSO sits behind a relatively undisturbed spiral arm (Figure reproduced from Keeney et~al. 2010).} 
\end{center}
\end{figure}

All indications are that this is a very strong interaction/merger in progress (note the tidal tail to the north and that the \ion{H}{II} regions are seen both in front of and behind the lenticular starlight component). So while the spiral arm/tidal tail in front of the quasar appears relatively undisturbed, its kinematic relationship to the gaseous spiral disk is uncertain; e.g., note that this outer spiral arm has a different pitch angle than the inner, more tightly-wound spiral arms. The UV and optical spectroscopy detects two absorption components at $cz = 5250$ \& 5500~\kms\ in various low-ionization metals including \ion{Mg}{II}, \ion{Fe}{II}, \ion{Ca}{II} and \ion{Na}{I} (Bergeron et~al. 1987; Keeney et~al. 2010). We also detect two \ion{H}{I} 21-cm absorption components at these same recession velocities (Figure~2). The strength of these two components predicts N$_{\rm H\,I} \geq 2\times10^{20}~{\rm cm^{-2}}$ for $T_{\rm spin} = 300$~K, which suggests that this overall system is a damped-\lya\ absorber (DLA), many of which have been discovered at high-$z$ (Wolfe, Gawiser, \& Prochaska 2005; Prochaska et~al. 2007). The lower velocity absorber is a likely LLS. However, we cannot confirm the DLA/LLS identifications without doubt because the wavelength of the associated \lya\ absorption is blocked by the presence of a higher redshift ($z=0.85$) LLS.

\begin{figure}[!t]
\begin{center}
\includegraphics[scale=0.35, angle=0]{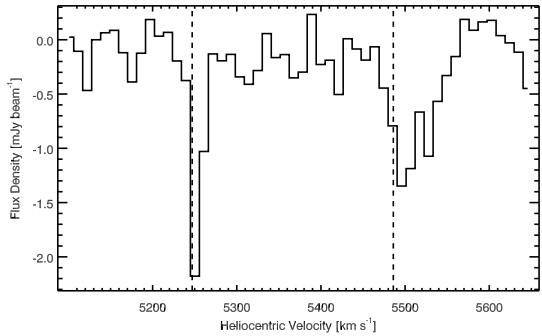}
\caption{\small The redshifted \ion{H}{I} 21-cm spectrum of the quasar PKS~1327--206 showing two components of absorption. The vertical dashed lines identify the velocities of the optical and UV metal-line absorptions from Bergeron et~al. (1987) and Keeney et~al. (2010). The faint \ion{H}{II} regions just NNW of the quasar have velocities similar to the broad \ion{H}{I} absorption (5500~\kms), and so we identify that \ion{H}{I} component with disk gas and the 5250~\kms\ \ion{H}{I} as an HVC.}
\end{center}
\end{figure}

The opportunity to observe a DLA very nearby offers additional insights provided by the high-resolution imaging, ground-based spectroscopy, and \ion{H}{I} VLA imaging spectroscopy only possible to obtain in this detail at very low-$z$. We identify the broad, and potentially multi-component, \ion{H}{I} 21-cm absorption line at 5500~\kms\ with disk gas associated with the outer spiral arm in ESO 1327--2041, which we detect at similar velocities in \ion{H}{I} emission. Additionally, faint \ion{H}{II} regions NNW of the quasar along the spiral arm (seen in Figure~1) have similar velocities to the 5500~\kms\ metals + \ion{H}{I} absorber, confirming our identification of this component as disk absorption. Therefore, we identify the 5250~\kms\ absorber with an HVC, but the kinematic status of the HVC gas is uncertain.  

While the orientation of the disk is established by the relationship of the \ion{H}{II} regions to the lenticular bar, the interaction/merger (and tidal tail stream to the north) suggest that significant anomalous gas motion is present in this system. So, while the galaxy orientation (SW edge is closer to us) and blue-shifted HVC gas suggest outflowing gas with respect to the underlying disk, this is far from certain since both the location (in front of or behind the disk from our point of view) and plane of motion of the HVC cloud relative to the disk gas is also uncertain. It seems quite unlikely that there are enough supernovae so far out in this spiral arm to drive a wind of several hundred \kms. Thus, a tidal origin for this high-velocity gas seems more plausible in this case. But we cannot make a strong case that all this HVC gas is infalling; evidently, both a fortuitous orientation and also a simple kinematic (i.e., not strongly interacting) system are required to make an unambiguous determination of an absorber's kinematics relative to a nearby galaxy. This is why the rare opportunity provided by the 3C~232/NGC~3067 system is so important to study in detail.

\section{3C~232/NGC~3067: An HVC Falling into a Nearby Galaxy} 

This low redshift system (Figure~3) consists of the very nearby SAB galaxy NGC~3067 ($cz=1465$~\kms) and the quasar 3C~232 ($z=0.53$) projected 11~kpc above the galaxy's nucleus. Due to the proximity of this bright quasar to the galaxy, this system has been the object of several studies over the years (Haschick \& Burke 1975; Boksenberg \& Sargent 1978; Stocke et~al. 1991; Carilli \& van Gorkom 1992; Tumlinson et~al. 1999; Keeney et~al. 2005). Attention to this system was first brought by Burbidge et~al. (1971) who used this and three other examples to argue for the non-cosmological redshift hypothesis. Detections of \ion{H}{I} 21-cm (Haschick \& Burke 1975), \ion{Ca}{II} (Boksenberg \& Sargent 1978) and \ion{Na}{I} (Stocke et~al. 1991) absorption at the redshift of NGC~3067 followed, including the discovery of multiple components in \ion{Na}{I} and \ion{Ca}{II} (Stocke et~al. 1991). The strongest of these three components is coincident with the \ion{H}{I} 21-cm absorber.

An \ion{H}{I} 21-cm VLA image (Carilli \& van Gorkom 1992) of this system shows an \ion{H}{I} emission cloud at the position of the quasar, a large diffuse halo of \ion{H}{I} surrounding the galaxy, and a discrete cloud $\sim\,$5~kpc in size located immediately in front of 3C~232. These VLA observations also provide an accurate \ion{H}{I} rotation curve for the galaxy, including velocities into the low halo. Arecibo, VLA and VLBA \ion{H}{I} absorption spectra in conjunction with a damped \lya\ profile from STIS yield a column density of $N_{\rm H\,I} = 1\times10^{20}~{\rm cm^{-2}}$ (making this absorber a LLS, nearly a DLA) and $T_{\rm spin}=500$~K, consistent with the gas kinetic temperature from the \ion{H}{I} 21-cm linewidth (Keeney et~al. 2005). Other \hst\ UV spectra detect three components in both low- and high-ionization lines including \ion{Si}{IV}, \ion{C}{IV}, \ion{Fe}{II}, \ion{Mn}{II}, \ion{Mg}{II} and \ion{Mg}{I} (Tumlinson et~al. 1999; Keeney et~al. 2005).  

A summary of the observable properties of the strongest absorber (the \ion{H}{I} 21-cm absorber) in this system made by Keeney et~al. (2005) finds a very striking parallel with the properties of HVCs around our own Galaxy. Specifically, the kinematics, size, temperature, gas density, location and metallicity of this very nearby LLS are consistent with this absorber being an HVC. The other two absorbers could also be HVCs, or possibly the more abundant, lower \ion{H}{I} column density highly-ionized HVCs (Sembach et~al. 2003; Richter et~al. 2009; Shull et~al. 2009). 

Tumlinson et~al. (1999) used the non-detection of the \ion{H}{I} 21-cm absorbing cloud (see also Stocke et~al. 1991) in \Ha\ emission and the absorption line strength ratios (particularly \ion{Fe}{II}/\ion{Fe}{I}) to set both upper and lower limits on the intensity of the  $z=0$ extra-galactic ionizing radiation field. This analysis also yielded a limit on the percentage of ionizing radiation escaping upwards from the disk to the location of this cloud ($<2\%$) consistent with other attempts to measure this quantity at low-$z$.  While the limits of Tumlinson et~al. (1999) and Stocke et~al. (1991) are larger than the expected \Ha\ surface brightness of a cloud irradiated by the extragalactic background (Bland-Hawthorn \& Maloney 1999), the discovery of \Ha\ emission associated with the Magellanic Stream has shown that the disruption of clouds moving through a hot galactic halo can enhance \Ha\ emission (Bland-Hawthorn et~al. 2007).  The absence of \Ha\ emission from the \ion{H}{I} cloud projected onto the quasar or between the galaxy and the quasar on the sky suggests that there is no strong outflowing starburst wind or shock heating due to an infalling cloud impacting a hot gaseous halo, and so this non-detection favors neither hypothesis.  But the question about whether these absorbers are infalling or outflowing depends critically upon the orientation on the sky of this galaxy since very accurate radial velocities are available from the \ion{H}{I} 21-cm emission and absorption. 

In Keeney et~al. (2005) we concluded that we are viewing the underside of NGC~3067 based upon the observed \ion{H}{I} rotation curve of Carlli \& van Gorkom (1992). This orientation implies that the blueshift of the \ion{H}{I} 21-cm absorber relative to the \ion{H}{I} systemic velocity of NGC~3067 means that the absorbing cloud is {\em infalling} onto the galaxy disk. The simplest model for these three absorbers, in which all total space velocities are perpendicular to the galaxy disk, finds the strongest system to be infalling at $-115$~\kms. The other two systems are one infalling and the other outflowing, although even for the latter the inferred outflow velocity is less than the escape velocity at its location above the disk. Thus, for this rather normal spiral galaxy there is no evidence for outflowing gas from the nucleus that escapes the galaxy's gravitational potential, and two of the three clouds are infalling gas being accreted onto the galaxy disk. By the perpendicular motion assumption, the third system will also return to the galaxy disk as a `galactic fountain'. Therefore, all the observed CGM gas detected in front of 3C~232 is a sampling of the total amount of CGM being accreted by NGC~3067 at this time or in the near future. 

Based upon similar UV absorption line spectroscopy with \hst/STIS and {\sl FUSE} of targets above and below our own Galactic Center, Keeney et~al. (2006a) come to similar conclusions for the Milky Way's Galactic wind; i.e., the outflowing gas on either side of the nucleus will not escape from the potential well of our Galaxy, rising to $\approx12.5$~kpc before returning to the disk like a fountain. These two examples are consistent with the Steidel (1995) hypothesis that LLS absorbers are the bound halos of luminous ($\sim {\rm L}^*$) galaxies. If this inference is correct, then these large galaxies are not the primary contributors to metals and energy in the low-$z$ IGM. We can also infer that LLSs are a sampling of the total amount of gas in the CGM that will eventually fall onto large galaxies to stimulate and enrich their star formation. Therefore, it seems worthwhile to confirm that the 3C~232/NGC~3067 LLS is infalling gas.

Figure~3 shows the \hst/WFPC2 {\em I}-band (F791W) image based on 2400~secs of observing time. The pure \Ha\ image shown in Figure~4 is based on 14\,400~secs of integration time in the F658N [\ion{N}{II}] filter minus the scaled {\em I}-band image. The F658N filter has a bandpass of 29~\AA\ FWHM centered at 6591~\AA\ and so contains only \Ha\ at the redshift of NGC~3067. The filter throughput is 79\% (its peak throughput) for \Ha\ throughout the velocity range of the disk of NGC~3067 seen in the \ion{H}{I} 21-cm images of Carlli \& van Gorkom (1992). The limiting fluxes for these two images are: (1) an {\em I}-band limiting magnitude (3$\sigma$) of $I_{\rm limit} = 26.0$ and (2) a 3$\sigma$ \Ha\ flux limit of $F({\rm \Ha})= 2 \times10^{-18}$ ${\rm ergs\,cm^{-2}\,s^{-1}}$. The latter limit corresponds to an \Ha\ luminosity of ${\rm L(\Ha)} = 1\times10^{35}~{\rm ergs\,s^{-1}}$ for an unresolved source ($<15$~pc) at the Hubble-flow distance to NGC~3067, which is about 0.3\% of the luminosity of the Orion Nebula, ${\rm L(\Ha)} \approx 3\times10^{37}~{\rm ergs\,s^{-1}}$. This limit corresponds to a star formation rate of $\sim 10^{-6}~{\rm M_{\odot}\,yr^{-1}}$ (Kennicutt 1998).

Also, there is no indication at all for \Ha\ emission knots between the disk of NGC~3067 and 3C~232 on the sky down to the 3$\sigma$ limit quoted above. The ground-based limits on diffuse \Ha\ from the \ion{H}{I} cloud in front of 3C~232 are exceptionally deep and place excellent limits on the flux of the extragalactic ionizing radiation at $z=0$ of $\Phi(\rm ion) \leq 80\,000~{\rm photons\,cm^{-2}\,s^{-1}}$ (Stocke et~al. 1991; Tumlinson et~al. 1999). But very small-scale emission features associated with extra-planar \ion{H}{II} regions or with gas collisionally-ionized by an outflowing wind from the disk could have been missed. However, we find no such emission knots in our pure \Ha\ image (Figure~4) so that there is no observational evidence supporting the hypothesis that the LLS cloud in front of 3C~232 was expelled by either a very active recent episode of star formation or an AGN, now dormant. If this were the case we would expect to see shock-heated gas above the plane of NGC~3067 on both sides of the nucleus as has been commonly observed in starburst galaxies (Veilleux, Cecil, \& Bland-Hawthorn 2005).

\begin{figure}[!t]
\begin{center}
\includegraphics[scale=0.2, angle=0]{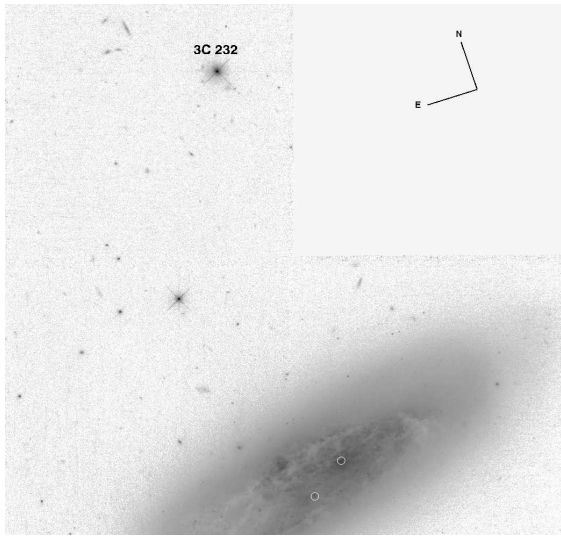}
\caption{\small \hst/WFPC2 {\em I}-band image of NGC~3067.  The position of the nearby quasar 3C~232 and the orientation of the image have been labelled.  The white circles (1~arcsec radius) mark the two possible nuclear positions (NW position preferred) discussed in the text.  The field of view of the image is approximately $150\times150$~arcsec, which corresponds to $15\times15$~kpc at the redshift of NGC~3067.}
\end{center}
\end{figure}

We can also use the pure \Ha\ image in Figure~4 to support the hypothesis that we are viewing NGC~3067 from underneath its disk; i.e. the southern edge is more distant from us. We have tested this hypothesis in two simple ways, by counting the \ion{H}{II} regions above and below the nucleus of the galaxy and by measuring the integrated \ion{H}{II} region \Ha\ flux on these two sides. We identify the nucleus of NGC~3067 using the dynamical center of the galaxy as inferred from the \ion{H}{I} 21-cm emission map (Carilli, van Gorkom \& Stocke 1989). Carilli et~al. do note that the dynamical location for the nucleus is well west of the apparent isophotal center suggesting that the galaxy's eastern edge has been truncated by a recent interaction.  While there is a large complex of \Ha\ emission and obscuration at that location (see Figure~3), there is no clear nuclear bulge visible to support this identification. The total {\em I}-band luminosity within 1~arcsec of that location is $(6.54\pm0.06)\times10^{7}~{\rm L_{\odot}}$. However, the only plausible alternative to this identification are the semi-stellar knot to the SW and its accompanying \ion{H}{II} region, which have a total {\em I}-band luminosity of $(2.29\pm0.02)\times10^{7}~{\rm L_{\odot}}$. These are also marked in Figure~3.

\begin{figure}[!t]
\begin{center}
\includegraphics[scale=0.2, angle=0]{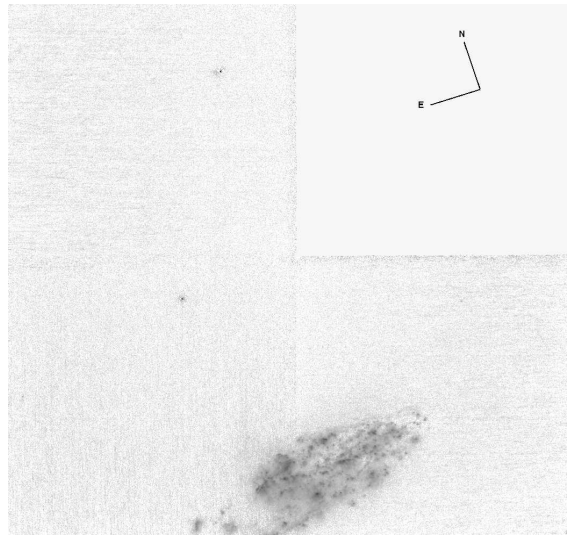}
\caption{\small \hst/WFPC2 \Ha\ image of NGC~3067.  The orientation of the image has been labelled. The faint dots north of NGC~3067 are the incomplete subtractions of the quasar and a bright field star (see Figure~3) and not \Ha\ emission.}
\end{center}
\end{figure}

Once the nucleus is identified, the interpretation of this image with respect to the galaxy orientation is straightforward; the more distant side suffers greater obscuration so that the numbers of \ion{H}{II} regions and their luminosities are diminished. We place a dividing line between the north and south sides through the nucleus along the major axis of the galaxy as defined by the outer ($>10$~arcsec from the nucleus) isophotes. The results of these measurements are that the ratio of \ion{H}{II} region numbers and integrated luminosities top/bottom are: $2.55\pm0.32$ and $1.44\pm0.01$, where the standard deviations are root-N in numbers and counts in the image. In these determinations we have used only the portion of the galaxy on either side of the major axis for which both the top and the bottom are fully on the CCD image. If we had used the small stellar emission knot to the SW (see above), these ratios would have been substantially larger. Both measurements clearly support the previous conclusion of Keeney et~al. (2005) based upon the \ion{H}{I} rotation curve as well as a visible inspection of Figures~3 \& 4, in which the entire south-west side of the galaxy disk appears truncated by a large-scale swath of obscuration most easily seen in Figure~3. We conclude that we are viewing the underside of NGC~3067 and that the LLS detected in optical/UV lines as well as \ion{H}{I} 21-cm absorption is infalling onto the galaxy disk. 

In addition, we use the red starlight in Figure~3 to calculate an improved axial ratio of the galaxy disk. Averaging the ellipticity for all isophotes with semi-major axes $>10$~arcsec from the nucleus yields an axial ratio of $0.35\pm0.01$, corresponding to a disk inclination of $73\pm1$~degrees, slightly larger than the $68^{+4}_{-3}$~degrees advocated by Rubin et~al. (1982) and used by us previously. The full spread of axial ratios seen in the outer disk covers a range of 12 degrees around our measured average, so there is no indication that the \hst\ image provides a much improved inclination angle for NGC~3067. 

In any case, the kinematic derivations for the absorbers do not change much at all. At this new inclination angle the derived velocities of the metal-line absorbers (assuming motion only perpendicular to the disk) are: $-330\pm40$, $-160\pm30$ and $+220\pm50$~\kms, somewhat larger than the values we derived earlier based upon the Rubin et~al. (1982) inclination value. This new inclination angle places the absorbers at a minimum height above the plane of NGC~3067 of $11.7\pm0.1$~kpc. Given the revised dynamically-determined mass of NGC~3067 of $(1.3\pm0.2)\times10^{11}~{\rm M_{\odot}}$ yields an escape velocity at 11.7~kpc of $310\pm25$~\kms. Our conclusion remains that given only perpendicular motion for these three clouds, even the outflowing CGM gas does not escape from NGC~3067. Relaxing the perpendicular motion assumption could allow this one gas cloud to escape from the galaxy for a fairly wide range of outflow angles, although at this height above the galaxy's disk we would expect any starburst or AGN wind to become collimated sufficiently so that the velocity vector is quite close to being purely vertical from the disk (e.g., MacLow \& McCray 1988).

\section{Discussion}

For many years it has been suggested that the ultimate promise of QSO absorption-line studies is to extend our detailed knowledge of local galaxies out to $z > 5$, where the highest redshift absorbers are found. The nearly flat selection function of absorption lines with redshift allows absorbers to be counted and their physical conditions assessed at whatever redshift distance they can be observed. In the current instance, the population of \ion{Mg}{II}/LLS absorbers has been studied in some detail by Bergeron \& Boiss\'{e} (1991), Steidel and Sargent (1992), Churchill et~al. (1999, 2000), Steidel et~al. (2002), Churchill, Vogt \& Charlton (2003), Kacprzak et~al. (2010) and others at $z=0.2$--2.0. For the current discussion we assume that the strong (rest-frame ${\mathcal W}_{\lambda} \geq 0.3$~\AA) \ion{Mg}{II} absorbers are synonymous with LLSs. 

In the large survey by Steidel \& Sargent (1992) DLAs comprise $<10\%$ of their strong \ion{Mg}{II} sample; the remainder are LLSs. Steidel (1995, 1998) found that 55 of his 58 absorbers with rest-frame Mg II equivalent widths $\geq0.3$~\AA\ in his sample at $0.2 \leq z \leq 1$ were associated with bright (${\rm L} \geq 0.1 {\rm L}^*$) galaxies at the same redshift within a projected distance of 50$\,h^{-1}$ kpc. The few absorbers without bright galaxy associations have been found to be DLAs, probably due to the sightline directly penetrating the disk of both luminous and dwarf galaxies so that the background QSO continuum makes the foreground galaxy hard to spot. The LLSs, being offset from the galaxy disk and being associated with luminous galaxies, means that the associated $\sim {\rm L}^*$ galaxies are much easier to detect. 

So, while the prima facie evidence is that LLSs are exclusively due to gas in the halos of luminous galaxies, McLin, Giroux \& Stocke (1998) used the Steidel (1995, 1998) statistic that 55 \ion{Mg}{II} absorbers in his sample are associated with $\geq 0.1 {\rm L}^*$ galaxies to conclude that LLSs could not be associated with dwarf galaxies {\bf at all}. Physically, the gas giving rise to a LLS is likely within a gravitationally-bound halo because the ionization energy of \ion{Mg}{II} is so close to that of \ion{H}{I} (15.0~eV compared with 13.6~eV). This means that any high column density \ion{Mg}{II} absorber must be successfully shielded from the extragalactic ionizing radiation field. An outflowing wind that smoothly varies in density (i.e., no clumps) will rapidly expand, lose its self-shielding and \ion{Mg}{II} ions will be quickly ionized to become \ion{Mg}{III}. This simple picture suggests that massive galaxies have bound gaseous halos and dwarf galaxies possess unbound winds, but these ideas need to be tested by observing other examples of LLSs and more highly-ionized metal-line systems associated with dwarfs and ${\rm L}^*$ galaxies at low-$z$ with \hst.

In the most recent work on \ion{Mg}{II}/LLSs, Kacprzak et~al. (2010) come to quite similar conclusions as we do here from the 3C~232/NGC~3067 system alone. Based on their \hst\ imaging and beautiful ground-based spectroscopy of $\sim {\rm L}^*$ galaxies nearby to low-$z$ LLSs (but at significantly higher redshifts than our prototype system NGC~3067), Kacprzak et~al. find: 
\begin{enumerate}
\item The \ion{Mg}{II} absorption almost always falls on one side or the other of the galaxy's velocity centroid. This requires that almost all of the \ion{Mg}{II}/LLS absorption is either infalling or outflowing and is rarely a combination of both.
\item These authors also show that the absorbing clouds seen as strong \ion{Mg}{II} absorbers are HVCs in the sense that their velocities exceed the velocity of the disk gas `beneath them' (see also, Steidel et al. 2002).
\end{enumerate} 
Further, the vector from the galaxy to the QSO (and thus to the absorbers) is rather randomly oriented with respect to the galaxy disk. It seems unlikely that an outflowing wind would be spherically symmetric around a large disk galaxy given the strong density gradients which orient an outflowing wind rather quickly into a perpendicular wind (e.g., MacLow \& McCray 1988). 

Therefore, we conclude that the Kacprzak et~al. results are suggestive that LLSs are dominated by infalling gas. However, lacking either detailed \ion{H}{I} 21-cm rotation curves or imaging with high enough physical resolution to determine galaxy orientation in the Kacprzak et~al. sample without doubt, it cannot be proven as yet whether the gas is infalling or outflowing in these cases. Despite these limitations, we note that the Kacprzak et~al. results are fully consistent with our interpretations of the nearby example of 3C~232/ NGC~3067.

It has been known for some time that the number of strong \ion{Mg}{II} absorbers per unit redshift is relatively constant with redshift ($dN/dz \approx 1$ per unit redshift; although the very strongest \ion{Mg}{II} absorbers do appear to exhibit significant cosmological evolution; Steidel \& Sargent 1992; Rao \& Turnshek 2000), requiring that the product of the number of sites for these absorbers, their area on the sky and their covering factor within that area does not vary significantly from $z=0.2$--1.0. If luminous galaxies are the sites for these absorbers, it is reasonable to expect that the numbers of sites for strong \ion{Mg}{II} systems would be relatively constant with look-back-time since these galaxies are already in-place by redshifts of one or higher. A constant number of strong \ion{Mg}{II} absorbers and a constant number of sites requires that the area on the sky $\times$ the covering factor of these absorbers is also constant over the time interval from $z=0$--1. This suggests that massive galaxy halos have seen little change in the last few Gyrs. 

Theoretical modelling of massive galaxy halos and cold gas clouds within them have the clouds forming either by cooling instabilities in a pre-existing hot halo (Kere\u{s} \& Hernquist 2009) or near the transition between an ionizing background dominated by extra-galactic sources and one dominated by local sources (Giroux \& Shull 1997). In either case, the radius at which these transitions/instabilities occur is not expected to change dramatically with redshift, in agreement with the \ion{Mg}{II}/LLS absorber statistics. But, what does change somewhat is the number of absorbers per LLS. 

\ion{Mg}{II} absorptions are so highly saturated that their equivalent width is increased not by an increasing optical depth but by an increased velocity spread within the absorption. This effect has been used by Churchill, Vogt \& Charlton (2003) to estimate the numbers of absorbers as a function of equivalent width. Although there is signficant scatter, these authors find that: $(N/17) \approx {\mathcal W}_{\lambda}$, where $N$ is the estimated number of individual absorption components within any one \ion{Mg}{II} absorption system and ${\mathcal W}_{\lambda}$ is the rest-frame equivalent width (in \AA) of the 2796~\AA\ \ion{Mg}{II} line. And the trend is for there to be more components in \ion{Mg}{II} absorbers at higher redshift by a factor of, $N \propto (1+z)^{0.6}$, not a very strong evolution at all (Nestor, Turnshek, \& Rao 2005). In the context of the results presented here for the LLS prototype 3C~232/NGC~3067, the size of the bound halo and the covering factor of the halo gas in NGC~3067 have not changed significantly but the number of HVC-like clouds either being accreted or as part of a galactic fountain has decreased modestly by a factor of at most 2--3 over the last few Gyrs. 

If most \ion{Mg}{II}/LLSs are composed of primarily infalling gas and bound galactic fountain gas like we find for the 3C~232/NGC~3067 system and for our own Galaxy, this helps to explain the \ion{Mg}{II} absorption line velocities seen by Kapcrzak et~al. (2010). This identification also suggests that \ion{Mg}{II}/LLSs are a sampling of the CMA gas rate falling onto these normal, $\sim {\rm L}^*$ galaxies. 

If this is the case, the available LLS statistics gathered by Bergeron \& Boiss\'{e} (1991), Steidel \& Sargent (1992), Rao \& Turnshek (2000), Churchill, Vogt \& Charlton (2003), and others can then be used to measure the amount of the CMA (although the simulators call this infalling gas `cold' because it is well below the virial temperature of the massive halo into which it is falling, observers call this `warm gas' because it has both a neutral and an ionized component). Extrapolated down to the current epoch, the \ion{Mg}{II} $dN/dz$ absorber densities account for only about 1/3 to 1/2 of the total CMA predicted by numerical simulations (e.g., Kere\u{s} et~al. 2009). 

So, if there is a connection between the observed LLSs and the simulated CMA, the total CMA infall rate must also include more highly-ionized gas than can be observed easily in \ion{Mg}{II}. This more highly-ionized gas recently has been shown to comprise a large fraction of the total accretion onto the Milky Way, $\sim 1~{\rm M}_{\odot}$ per year in ionized gas as traced by \ion{Si}{III} (Shull et~al. 2009) and other ions (Richter et~al. 2009). Furthermore, the predicted increase of the CMA with redshift, i.e., $\dot{\rm M} \propto (1+z)^{2.5}$ (Kere\u{s} et al. 2009), is a much steeper dependence on redshift than the observed \ion{Mg}{II} statistics alone support (see above).

We conclude that the ionized fraction of the CMA must increase rapidly with redshift in order to bring the observations and the results of numerical simulations into even rough agreement. This seems quite reasonable since the intensity of the ambient ionizing radiation field around these clouds is increasing with redshift at a similar rate to the predicted CMA rate. This is true regardless of whether the ambient ionizing radiation is due to the extragalactic ionizing flux or the ionizing flux leaking upward from young stars in the nearby galaxy disk. Given a constant leakage factor ($f_{\rm esc}$) with redshift, the ionizing flux leaking up from the disk is also expected to increase at a similarly rapid rate [i.e., $(1+z)^{2.5}$] because the cosmic star formation rate also increases dramatically between $z=0$--1. This more highly-ionized CMA can be found only by detecting species like \ion{Si}{III}, \ion{Si}{IV}, \ion{C}{IV} and \ion{O}{VI} which are at UV wavelengths for absorbers at $z \leq 1.4$. Thus, UV spectroscopy with \hst\ is required.

\section{The Future is Now}

If big galaxies do not provide most of the metals to enrich the IGM at low-$z$, then dwarf galaxies must be the primary contributors. There is some, but not overwhelming, support for this hypothesis (Stocke et~al. 2004; Keeney et~al. 2006b). But, there also is significant evidence that the much more massive `Lyman-break galaxies' at $z\approx2$--4 have enormous starburst winds that escape into the IGM where nearby metal-line absorbers have been found (Adelberger et~al. 2003). So we expect any conclusions involving the source of IGM metals to remain controverisal and a topic for on-going research. Likewise the types of galaxies which possess bound gaseous halos will also remain controversial. While large-aperture, ground-based spectroscopy can address questions of IGM enrichment and galaxy halos and winds at $z\geq2$ directly, only \hst\ UV spectroscopy can study the low-$z$ IGM and its gas suppliers.

\begin{figure}[!t]
\begin{center}
\includegraphics[scale=0.4, angle=0]{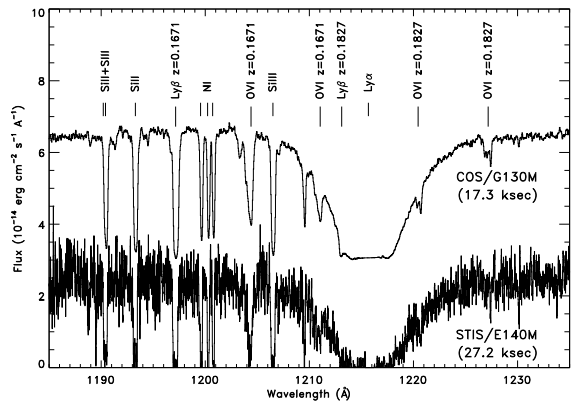}
\caption{\small A comparison between the STIS E140M echelle spectrum (${\rm SNR} \sim 12$) of PKS 0405--123 (bottom spectrum) and a shorter exposure of the same target with COS (${\rm SNR} \sim 50$) on top.  The two spectra have been offset in the y-coordinate for display purposes.  The COS spectrum is a portion of the `Early Release Observations' taken this past summer.  GTO and GO observations with COS are now well underway.}
\end{center}
\end{figure}

This is where the road ahead is quite bright due to the successful deployment of COS on \hst\ this last May. In \hst's Cycle 17 alone there will be $>$ 300 \hst\ orbits of COS observations made by 5 separate proposers (COS GTO Team $+$ 4 GOs) investigating QSO/galaxy pairs to study gas infall and outflow in and around a diverse set of low-$z$ galaxies. These chance projections of QSOs near foreground galaxies will probe the physical condition and kinematics of gas around M~31, around nearby and distant  starburst galaxies, and around normal luminous and dwarf galaxies in the local Universe.  We expect that these new observations will revolutionize this field and give us a much clearer picture of galaxy/gas interactions in the local Universe. We also expect that these and other COS observations will allow the assertions made in this paper based on the observational evidence of a very small number of low-$z$ absorbers to be thoroughly tested.
 
As a preview of what is to come, Figure~5 shows a pair of spectra of the same target, the bright $z=0.573$ quasar PKS~0405--123. The STIS spectrum has a resolution of 35\,000 and a ${\rm SNR} \sim 12$ based on 27~ksecs of observation. The COS spectrum has $R=18\,000$ and ${\rm SNR} \sim 50$ based on 17~ksecs of observation. Based upon the nominal in-orbit performance of the COS instrument, the next few years should see a substantial change in our understanding of the local IGM, wherein most of the baryons are still lurking. Interesting topics in local IGM studies which will be addressed by COS observations include: which types of galaxies possess outflowing winds that escape into
the IGM, the extent of the spread of metals away from galaxies into the IGM, the metallicity of gas in cosmic voids and the amount of baryons in the shock-heated `warm-hot IGM' (WHIM). The present authors look forward to using \hst/COS to contribute to our better understanding of these topics.

\section*{Acknowledgments} The authors acknowledge the support of an \hst\ guest observing grant GO-10925.01-A and the Cosmic Origins Spectrograph project through NASA's Goddard Space Flight Center (grant \#NNX08AC14G) and the Space Telescope Science Institute.

\section*{References}
% References are listed as in the following example, for more examples, please
% see the PASA Style Guide
\reference Adelberger, K. L., Steidel, C. C., Shapley, A. E., \& Pettini, M. 2003, ApJ, 584, 45
\reference Bergeron, J., D'Odorico, S., \& Kunth, D. 1987, A\&A, 180, 1
\reference Bergeron, J. \& Boiss\'{e}, P. 1991, A\&A, 243, 344
\reference Bland-Hawthorn, J., \& Maloney, P. R. 1999, ApJL, 510, L33
\reference Bland-Hawthorn, J., Sutherland, R., Agertz, O., \& Moore, B. 2007, ApJL, 670, L109
\reference Boksenberg, A. \& Sargent, W. L. W. 1978, ApJ, 220, 42
\reference Burbidge, E. M., Burbidge, G. R., Solomon, P. M., \& Strittmatter, P. A. 1971, ApJ, 170, 233
\reference Carilli, C. L., van~Gorkom, J. H., \& Stocke, J. T. 1989, Nature, 338, 134
\reference Carilli, C. L. \& van~Gorkom, J. H. 1992, ApJ, 399, 373
\reference Churchill, C. W., Rigby, J. R., Charlton, J. C., \& Vogt, S. S. 1999, ApJS, 120, 51
\reference Churchill, C. W. Mellon, R. R., Charlton, J. C., Jannuzi, B. T., Kirhakos, S., Steidel, C. C., \& Schneider, D. P. 2000, ApJ, 543, 577
\reference Churchill, C. W., Vogt, S. S., \& Charlton, J. C. 2003, AJ, 125, 98
\reference Froning, C. S. \& Green, J. C. 2009, Ap\&SS, 320, 181
\reference Giroux, M. L. \& Shull, J. M. 1997, AJ, 113, 1505
\reference Haschick, A. D. \& Burke, B. F. 1975, ApJL, 200, L137
\reference Kacprzak, G., et~al. 2010, ApJ, submitted 
\reference Keeney, B. A., Momjian, E., Stocke, J. T., Carilli, C. L., \& Tumlinson, J. 2005, ApJ, 622, 267
\reference Keeney, B. A., Danforth, C. W., Stocke, J. T., Penton, S. V., Shull, J. M., \& Sembach, K. R. 2006a, ApJ, 646, 951
\reference Keeney, B. A., Stocke, J. T., Rosenberg, J. L., Tumlinson, J., \& York, D. G. 2006b, AJ, 132, 2496
\reference Keeney, B. A., Stocke, J. T., Danforth, C. W., \& Carilli, C. L. 2010, AJ, submitted 
\reference Kennicutt, R. C., Jr. 1998, ARA\&A, 36, 189
\reference Kere\u{s}, D., Katz, N., Fardal, M., Dav\'{e}, R., \& Weinberg, D. H. 2009, MNRAS, 395, 160 
\reference Kere\u{s}, D. \& Hernquist, L. 2009, ApJL, 700, L1
\reference MacLow, M.-M. \& McCray, R. 1988, ApJ, 324, 776
\reference McLin, K., Giroux, M. L., \& Stocke, J. T. 1998, in ASP Conf. Ser. 36: Galactic Halos, ed. D.~Zaritsky (Provo: ASP), 175
\reference Nestor, D. B., Turnshek, D. A., \& Rao, S. M. 2005, ApJ, 628, 637
\reference Prochaska, J. X., Wolfe, A. M., Howk, J. C., Gawiser, E., Burles, S. M., \& Cooke, J. 2007, ApJS, 171, 29
\reference Rao, S. M. \& Turnshek, D. A. 2000, ApJS, 130, 1
\reference Richter, P., Charlton, J. C., Fangano, A. P. M., Bekhti, N. B., \& Masiero, J. R. 2009, ApJ, 695, 1631
\reference Rubin, V. C. Ford, W. K., Jr., Thonnard, N., \& Burstein, D. 1982, ApJ, 261, 439
\reference Sembach, K. R., et~al. 2003, ApJS, 146, 165
\reference Shull, J. M., Jones, J. R., Danforth, C. W., \& Collins, J. A. 2009, ApJ, 699, 754 
\reference Steidel, C. C. 1995, in QSO Absorption Lines, ed. G.~Meylan (Garching: Springer), 139 
\reference Steidel, C. C. 1998, in ASP Conf. Ser. 36: Galactic Halos, ed. D.~Zaritsky (Provo: ASP), 167
\reference Steidel, C. C. \& Sargent, W. L. W. 1992, ApJS, 80, 1
\reference Steidel, C. C., Kollmeier, J. A., Shapley, A. E., Churchill, C. W., Dickinson, M., \& Pettini, M. 2002, ApJ, 570, 526
\reference Stocke, J. T., Case, J., Donahue, M., Shull, J. M., \& Snow, T. P.  1991, ApJ, 374, 72
\reference Stocke, J. T., Keeney, B. A., McLin, K. M., Rosenberg, J. L., Weymann, R. J., \& Giroux, M. L. 2004, ApJ, 609, 94
\reference Tumlinson, J., Giroux, M. L., Shull, J. M., \& Stocke, J. T. 1999, AJ, 118, 2148
\reference Veilleux, S., Cecil, G., \& Bland-Hawthorn, J. 1995, ApJ, 445, 152
\reference Whitmore, B. C., Lucas, R. A., McElroy, D. B., Steinman-Cameron, T. Y., Sackett, P. D., \& Olling, R. P. 1990, AJ, 100, 1489
\reference Wolfe, A. M., Gawiser, E. \& Prochaska, J. X. 2005, ARA\&A, 43, 861

\end{document}